\documentclass[pra,aps,showpacs,a4paper]{revtex4}
\usepackage[matrix,frame,arrow]{xypic}
\input{Qcircuit}
\usepackage[hmargin=0.7in,vmargin=1in]{geometry}
\usepackage{graphics}
\usepackage{times,mathptm}
\usepackage{epsfig}
\usepackage{mathrsfs}
\usepackage{amsmath,amssymb}
\usepackage[colorlinks]{hyperref}
\usepackage{subfigure}

\newcommand{\beq}{\begin{equation}}
\newcommand{\eneq}{\end{equation}}
\newcommand{\beqy}{\begin{eqnarray}}
\newcommand{\eneqy}{\end{eqnarray}}

\begin{document}

\title{On fault-tolerance with noisy and slow measurement \& preparation}
\author {Gerardo~A. Paz-Silva}
\altaffiliation{E-mail: {\tt gerardo.paz-silva@mq.edu.au
}}
\author {Gavin K. Brennen}
\author {Jason Twamley}

\affiliation{Centre for Quantum Computer Technology, Macquarie University, Sydney, NSW 2109, Australia}

\date{\today}

\begin{abstract}
{
It is not so well-known that measurement-free quantum error correction protocols can be designed to achieve fault-tolerant quantum computing. Despite the potential advantages of using such protocols in terms of the relaxation of accuracy, speed and addressing requirements on the measurement process, they have usually been overlooked because they are expected to yield a very bad threshold as compared to error correction protocols which use measurements. Here we show that this is not the case.  We design fault-tolerant circuits for the 9 qubit Bacon-Shor code and find a threshold for gates and preparation of $p_{(p,g) thresh}=3.76 \times 10^{-5}$ ($30\%$ of the best known result for the same code using measurement based error correction) while admitting up to $1/3$ error rates for measurements and allocating no constraints on measurement speed. We further show that demanding gate error rates sufficiently below the threshold one can improve the preparation threshold to $p_{(p)thresh} = 1/3$. We also show how these techniques can be adapted to other Calderbank-Shor-Steane codes.}      
\end{abstract}


\maketitle

An ideal quantum computer is a theoretical object capable of highly  
efficient computation. A major difficulty with the realization of such a powerful theoretical object  
is that physical implementations of any quantum operation will  
be noisy. However, with the use of quantum error correction (QEC) codes, fault-tolerantly  
designed circuits, and provided that error rates are below some threshold  
value, one is still able to efficiently simulate a quantum computation  
with arbitrarily high accuracy~\cite{Aharonov:99, Gottesman, Aliferis}. Experimental state of the art results show that error rates and execution times required for operations in order to achieve the fault-tolerant regime are not currently available. The results in this paper will alleviate part of this constraint pushing required error rates a step closer to current technology.

In many physical systems measurements pose a potential bottleneck for scalable fault-tolerant quantum computation because they are slower and/or noisier than gates or preparation~\cite{ITrap,Meunier:06}. However, they are central in the readout stage, and are widely used in QEC routines as a way of extracting error syndrome information in order to correct the quantum data. Slow measurements have been shown to be a surmountable issue by using error correction where measured error syndromes can be classically post-processed at the end of a round of gates to execute a compensating Pauli frame rotation~\cite{Aliferis&DV}, with the caveat that there can be a significant time lag during classical processing~\cite{Knill:05}. 
Regarding noise, measurement error rates cannot usually be  
improved by noise suppression techniques, i.e. dynamical decoupling,  
whereas gates can be~\cite{LidarDD, NewLidar}. Furthermore, measurement results  
must be $distinguishable$ in every time step, i.e. one must be able to discriminate  
between results from different measurements repeatedly over the computation, which leads to further constraints on the physical processes executing the measurements, e.g. measurements relying on  
photon scattering as in ion traps.~\cite{Myerson:08}

{ In this letter we overcome these problems by eliminating most measurements during fault-tolerant computation. It is well-known~\cite{Aharonov:99, Ngate} that this is possible for Calderbank-Shor-Steane codes~\cite{CSS} such as the Steane code, however {\it `` The penalty paid in the stringency of the threshold has never been quantified, but it is expected that replacing measurement by coherent operations decreases the noise threshold by a large amount"}\cite{Aliferis&DV}. We show that contrary to these conjectures coherent FT QEC suffers only slightly in regards to the threshold and brings substantial rewards. } 
 
We begin by setting up our scenario and introducing measurement free error correction (EC) routines for the Bacon-Shor code. We then show how to execute fault-tolerant Clifford operations consisting of: (I) preparation of $\ket{0}$ and $\ket{+}=(\ket{0}+\ket{1})/\sqrt{2}$ states, (II) Clifford group~\cite{fn01} unitary gates, and (III) measurement in the $X$ and $Z$ basis, and derive a threshold error rate which is stringent for preparation and gates but as high as 33\% for measurement. We proceed to show that through an encoder circuit we can prepare special ancillas at any level of concatenation. While this encoding cannot achieve an arbitrary low error rate ($p_{anc}$), it is small enough, $p_{(anc)} < p_{H-anc} =\sin^2(\pi/8)\sim 14.6\%$, to be used as a resource in magic state distillation~\cite{Knill:05, Bravyi:05}(MSD), a protocol using exclusively Clifford operations to distill arbitrarily low-error encoded $\ket{H_L}= (\ket{0_L} + e^{i \pi/4}\ket{1_L})/\sqrt{2}$ $magic$ states.  Using this resource to execute non-Clifford gates at the highest level of concatenation completes the universality of our model. Moreover, we show how to relax the threshold value for preparation, using a variant of algorithmic cooling and demanding a gate error rate, $p_{(g)}$, sufficiently below the threshold. { Thus fault tolerant universal quantum computing (FTUQC) can be achieved with measurement and preparation error rates, $p_{(p)}$ and $p_{(m)}$ respectively, which are already within reach of current technology. }

We demonstrate our scheme for the 9-qubit Bacon-Shor (BS) subsystem code~\cite{Bacon:06} but our tools can be adapted to other CSS codes (see~\ref{AppB}). The BS code is defined by the stabilizer set on a two dimensional array, 
\beq
S = \Bigg\{ \begin{array}{ccc} X&X&I\\X&X&I\\X&X&I  
\end{array}, \begin{array}{ccc} I&X&X\\I&X&X\\I&X&X \end{array},  
\begin{array}{ccc} Z&Z&Z\\Z&Z&Z\\I&I&I \end{array},\begin{array}{ccc}  
I&I&I\\Z&Z&Z\\Z&Z&Z \end{array} \Bigg\}.
\eneq
For this code logical Pauli operators are given by $X_L =  
\prod_{i=1}^{3} X_{i,1}; Z_L = \prod_{i=1}^{3} Z_{1,i}$ modulo  
stabilizer operations, i.e. $X_L$ ($Z_L$) acts on a column (row) of the array. 
This code is a subsystem code and is invariant under pairs of $X(Z)$ operators along any given row(column) because they act only on gauge degrees of freedom. Given the subsystem structure of the code one is able to correct acting on only one row (for X-errors) and one column (for Z-errors). The library of physical (level-0) gates we use is $\{ X, Z, H, CNOT, TOFFOLI, Z-TOFFOLI =  H^{\otimes 3} (TOFFOLI) H^{\otimes 3}, \ket{0} \textrm{ preparation},   
\ket{+}\textrm{ preparation},\ket{H}$\\$\textrm{preparation}$, $Z\textrm{-measurement}, X\textrm{-measurement}\}$, allowing also for 
non-local interactions.  We adopt an adversarial, local,  
stochastic error model~\cite{Error}.

\begin{figure}[h]
\subfigure[]{
\includegraphics[width=0.4 \columnwidth]{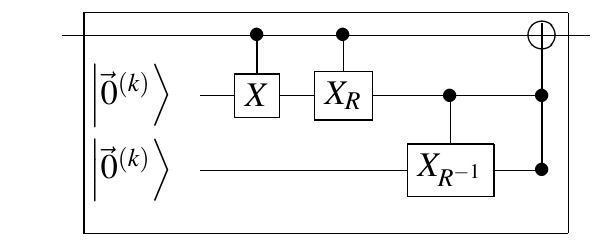}       
\label{M-gate}
}
\subfigure[]{
\includegraphics[width=0.4 \columnwidth ]{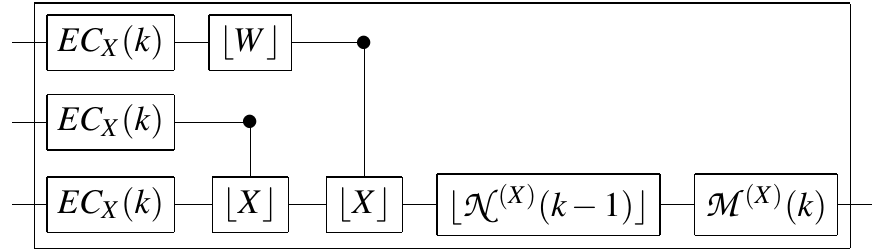}
\label{VN}}
\subfigure[]{
\includegraphics[width=0.8\columnwidth]{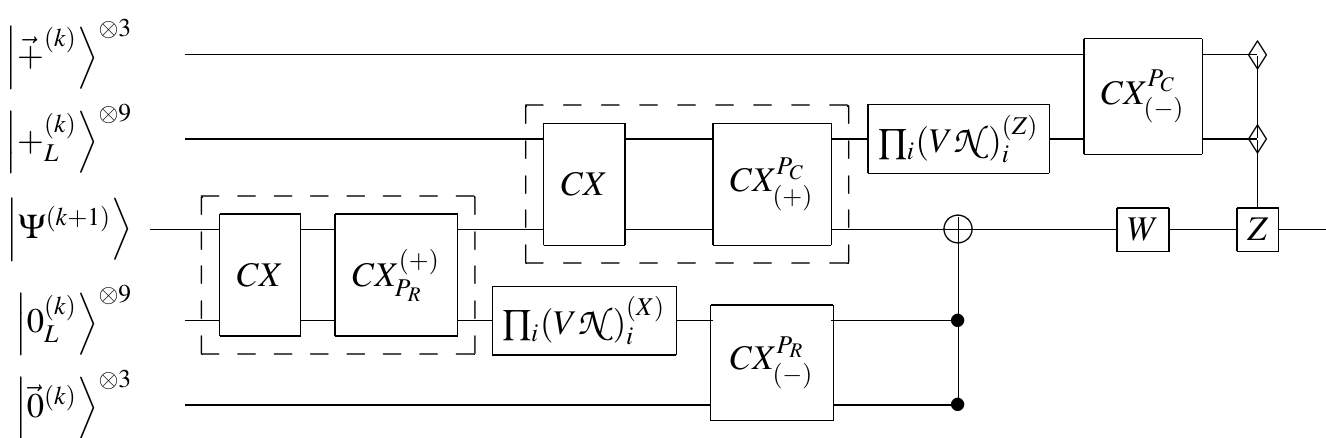}
\label{BS}}

\caption{ Measurement-free QEC routines for the QR and BS code. The inputs are $\ket{\vec 0^{(k)}}= \ket{000}^{\otimes 3^{k-1}}$ and $\ket{\vec +^{(k)}}= \ket{+++}^{\otimes 3^{k-1}}$.  
\textit{(a) The $\mathcal{M}$ gate.} An X-encoded majority voting gadget of level-$(k+1)$ of concatenation. Here all CNOTs are bitwise, i.e. each CNOT depicted corresponds to three $CNOT(k)$, and subscript $R$ corresponds to a cyclic $k$-encoded rotation of the targets of the corresponding gate. In the QR code the TOFFOLI gate depicted is bitwise. The $\mathcal{M}$ gate can also  
be designed for a Z-encoded quantum majority voting, with $\ket{\vec{+}^{(k)}}$  
ancillas and the obvious Hadamard conjugation of gates. When the need  
to distinguish them arises we shall denote $X$ and $Z$ encoded  
majority votings $\mathcal{M}^{(X)}$ and $\mathcal{M}^{(Z)}$  
respectively.
\textit{(b)} A subroutine acting on ancilla for processing error syndrome information extracted from the data.  The circuit shows one row, $(V\mathcal{N})_i (k)$, of the fully contracted exRec $V\mathcal{N}(k)=\lfloor \mathcal{N}(k+1)\rfloor $ representing a collection of $k$-level protected gates acting on row $i$ of ancilla which take part in an $EC(k+1)$ step. Note that in this circuit the output top lines are discarded so no EC gadget must protect them. With this, the exRec corresponding to $\mathcal{N}$ at degree of concatenation $k$ is $\mathcal{N}(k) = EC(k) \times \prod_{i \in rows} (V\mathcal{N})_i (k-1) \times EC(k)$. In our circuits $\lfloor G(k) \rfloor$ denotes the implementation of gate $G$, in terms of level-$(k-1)$ gates, without the prepended and appended EC(k) routines, and $W$ denotes a waiting gate.     
\textit{(c) Full error correction (EC) gadget for the BS code.} Here, a  
TOFFOLI with $\diamond$ controls is a $Z-TOFFOLI$; $CX =  
\prod^{3}_{i,j=1} CNOT^{(k)}_{(c,i,j),(t,i,j)}$ is a set of transversal  
CNOTs, $CX^{(\pm)}_{P_R} = \prod^{3}_{i,j=1} CNOT^{(k)}_{(c,i,j),(t,i,j \pm 1)}$
and $CX_{(\pm)}^{P_C} = \prod^{3}_{i,j=1} CNOT^{(k)}_{(c,i,j\pm 1),(t,i,j)}$. The control of the gates in boxes is always the top input of the gate. The last gate is a bTOFFOLI.}
\end{figure}

The first obstacle is of course to design a EC routine/gadget which  
uses coherent feedback instead of measurements \& feedback. One needs to use more gates within the EC gadgets to execute the coherent feedback and, in particular, one would typically need  
fault-tolerant implementations of TOFFOLI gates at every level. This would 
yield a very bad threshold value~\cite{Aharonov:99, Aliferis&DV}. However, during QEC we do not really need a full-fledged TOFFOLI gate since it will only be controlled by ancillas containing the syndrome, i.e. classical, information. For instance when correcting X-errors, a Z error in the ancillas is irrelevant, thus we can map a BS encoded ancilla to a quantum repetition (QR), i.e. bit-flip, code which protects against X errors but that is vulnerable to Z-errors. Using the QR encoded controls, and the structure of the logical operators in the BS code, we can use bitwise TOFFOLI gates to implement the needed operation (see Fig. {\eqref{M-gate}}). 

{
The mapping between the BS code and the QR code, of the same level of concatenation, is achieved using the gate $\mathcal{N}(k)$ $\mathcal{N}(k): \ket{s^{(k)}_L} \rightarrow {\ket{ \overrightarrow{(s)}^{(k)}}}$,
where $\ket{s^{(k)}_L}$ is encoded in the $k$-concatenated BS code ( $9^{k}$ physical qubits), and $ \ket{\vec{s}^{\;(k)}}$ denotes a bit encoded in $k$-concatenated QR code ($3^{k}$ physical qubits). From our joint use of BS and QR codes we must also introduce an error correction measurement-free routine for the QR code, i.e. states of the form $a \ket{0,0,0} + b\ket{1,1,1}$. We build a majority voting gadget, which we dub the $\mathcal{M}$-gate (Fig.\eqref{M-gate}). In the QR code all gates involved in the $\mathcal{M}$ gate are transversal and thus we can use this circuit as an EC gadget for this code at any level of concatenation. Moreover, through the $\mathcal{N}$ gate we can also use $\mathcal{M}$ as an encoded majority voting gadget, i.e. acting on a state of the form $a \ket{0^{(k)}_L, 0^{(k)}_L,0^{(k)}_L} + b\ket{1^{(k)}_L,1^{(k)}_L,1^{(k)}_L}$. By virtue of the fact that the Bacon-Shor code is, in essence, a composition of X and Z basis QR codes, we can use $\mathcal{M} \& \mathcal{N}$ as the building block for the BS EC gadget. }

Schematically the BS QEC routine works as follows (we refer the reader to Fig.~\eqref{BS} for a detailed description). The boxed part of Fig.~\eqref{M-gate} is a syndrome extraction stage, and turns the ancilla, initially in a $\vec{0}$ state, into a string which contains the error information. We adapt this method to the BS code. In this code, to correct for X-errors, we execute an extraction stage in every column of the BS state and get three strings $(s_1,s_2,s_3)$. We use them to vote into a fourth one $s_4 = s_1 \oplus s_2 \oplus s_3$, which will control final correction via $\mathcal{N}$ and bTOFFOLI. A single error in e.g. column one of the BS state leads to $s_4 = s_1$ which would correctly execute the correction by virtue of the gauge freedom; on the other hand a gauge operation, e.g. two $X$-errors in the same row, leads to $s_4 = s \oplus s = \vec{0}$ which correctly implies an identity correction operation. An analogous  analysis holds for $Z$-error correction.

Now the X and Z correction stages of the BS QEC routine are essentially equivalent but have some differences. { Because the syndrome information after the syndrome extraction stage is different in both cases, we define $\mathcal{N}^{(X)}$ (and $\mathcal{N}^{(Z)}$), gates for the BS code  
at level $k$ of concatenation (see Fig.\eqref{VN}): $\mathcal{N}^{(X)}(k) \equiv \prod_{i\in rows} (V  \mathcal{N}^{(X)})_i(k-1) = \prod_{i} CNOT_{(1,i),(3,i)}(k-1) \prod_{i} CNOT_{(2,i),(3,i)}(k-1) \prod_j \mathcal{N}^{(X)}_{(j,3)}(k-1)$, $\mathcal{N}^{(Z)}(k) \equiv \prod_{j\in columns} (V  \mathcal{N}^{(Z)})_j(k-1) = \prod_{i} CNOT_{(i,3),(1,1)}(k-1) \prod_{i} CNOT_{(i,3),(i,2)}(k-1) \prod_j \mathcal{N}^{(Z)}_{(3,j)}(k-1)$, where $A_{(r,c)}$ denotes gate A acting on the qubit in row $r$ and column $c$ of the logical qubit. The $(X)$ or $(Z)$ version of the gate is chosen depending on the correction subroutine in which it is being used, e.g. to correct X-errors (as in the lower part of Fig.~\eqref{BS}), we use $\mathcal{N}^{(X)}$. The $\mathcal{N}^{(X)}(\mathcal{N}^{(Z)})$ is a Z(X) decoder, where one keeps only the convenient protection while completely unprotectng against the other type of errors. Moreover, after the X syndrome extraction stage, the corresponding ancilla does not need protection against Z error, so only the lower stage ($EC_X$) of EC must be used. This greatly reduces the overall execution time for encoded gates acting on those ancillas. We found that due to this property, the subroutine $(V \mathcal{N})_i(k)$ not only takes less time, in terms of execution time of level-$(k-1)$ gates, but it can be shown to fail with a probability smaller than a CNOT(k), for $k > 1$. For $k=1$ EC gadgets, there is no need to use $\mathcal{N}$ since $\mathcal{N}(0) = Id$. We detail this in the Supplementary Material.\ref{AppA}

We are now ready to describe the remaining elements of our BS code fault-tolerant scheme. First we describe the elements needed to fault-tolerantly simulate any circuit based solely on Clifford operations.
{\it (I) Preparation of $\ket{0_L}$ and $\ket{+_L}$states:} by (i) starting  
with a $3\times3$ array of $\ket{+}$, and (ii) applying a $\mathcal{M}^{(X)}$ in every column we can prepare a $\ket{+_L}$. Similarly $\ket{0_L}$ is obtained by (i) starting with a $3\times3$ array  
of $\ket{0}$, and (ii) executing a $\mathcal{M}^{(Z)}$ in every row. {\it (II) Clifford group generators: $CNOT, H, Z^{1/2}$:} The CNOT gate is transversal, the $H$ gate can also be implemented in a bitwise fashion but, because stabilizers are rotated by this action, it is followed up by a $physical$ $\pi/2$-rotation accommodated by relabeling or rewiring of gates. The $Z^{1/2}$ gate can be implemented using  
the circuit in Fig.~\eqref{zhalf}, provided one can  
prepare a logical ancilla in $\ket{\pm i_L} = (\ket{0_L} \pm i \ket{1_L})/\sqrt{2}$.  
Since the $Z^{1/2}$ gate is not part of the EC routines, it is only needed at the highest level of concatenation. Furthermore, as it is the only complex gate, it can be shown that by always using the same logical ancilla prepared in $\ket{0_L} =1/\sqrt{2}(\ket{+i_L} + \ket{-i_L})$ to activate the circuit in Fig.~\eqref{zhalf}, then the entire quantum computation splits into two noninterfering paths (evolution by $U_{comp}$ and $U_{comp}^{\ast}$) and the measurements of real, Hermitian operators at the end have the same expectation values as for evolution by $U_{comp}$ alone~\cite{Dennis:02}.  Alternatively one can use the distillation circuit in \cite{Aliferis} at the highest level provided one can prepare it with an error rate below $p_{(i-anc)} = 1/2$.
{\it (III) X and Z basis measurements.-} They are only required at the  
highest level of concatenation. Given their form, measuring encoded logical operators can be achieved measuring only one row or column of the $9^{k} \times 9^{k}$ encoding array.\\
{\it Threshold calculation for Clifford operations.-} We use the  
extended rectangle (exRec) method developed in~\cite{Aliferis} to compute the threshold (see Supplementary Material for more details). An exRec of a gate is constructed by prepending and appending error correction routines on the inputs and outputs. The exRec with the largest number of malignant pairs, i.e. the number of pairs of faults which generate two or more errors in the data, will determine the threshold value. A quick inspection reveals that the largest exRec is the one corresponding to the CNOT gate. Following~\cite{Aliferis}, only at level $k=1$ must one consider all elements: preparation and gates (including waiting gates). At level $k>1$, using contraction of exRecs, preparation  
locations can be omitted. {This means that one has to solve the  
recursion relationships for the error $p^{(j)}$ at level $j$:
\beq
\label{recur}
p^{(1)} \leq A'_{(k=1)} (p^{(0)})^2;\quad p^{(k)} \leq A'_{(k>1)} (p^{(k-1)})^2,\textrm{  for  } k>1,
\eneq
where $A'_{(k)} = \frac{A_{(k)}}{2} \left(1 + \sqrt {1 + \frac{4 B}{(A_{(k)})^2}}\right)$, $B$ denotes all possible three-site errors, and $A_{(k)}$ denotes the number of malignant pairs in the largest exRec of that level. This process can be repeated for four site errors, etc. to get an even tighter bound~\cite{Aliferis}. Executing this algorithm with our largest exRec, the CNOT, we obtain a threshold value, for preparation and gates, $p_{(p,g)thresh} = 3.76 \times 10^{-5}$. }This value is not a bound for measurement error rates since they are not needed during the QEC process and are only required at the highest level of concatenation. So it follows that
\beqy
\label{meas} p_{(m)}^{(k+1)} &\leq& 3 (p_{(m)}^{(k)})^2 + \mathcal{O} (p^{(k)}). 
\eneqy
If preparation and gate error rates are below threshold, then for $k$ large enough $p^{(k)}$ is vanishingly small and the terms $\mathcal{O} (p^{(k)})$ can be neglected. Then the threshold condition for $X$ and $Z$ measurements is $p_{(m)thresh} = 1/3$.

{\it Encoded non-Clifford operations.-} The missing component to  
achieve universality is the FT execution of a non-Clifford gate. Using the circuit in Fig.~\eqref{pi4} we translate the problem into  preparing the $\ket{H_L}$ ancilla. To create an ancilla at the highest level
we will use an encoder circuit which will allow us to keep the $p_{(m)thresh} \leq 1/3 $. To encode an arbitrary state we use the following algorithm: (i) we start with the level-0 state $\ket{\phi}$ we want to encode and 8 $\ket{0}$ states, then (ii) we use CNOT gates, including waiting times such that never in one step does one qubit interact with more than one qubit, to create the state  $\ket{\vec \phi}_{3\times3}= a\ket{\vec 0}_{3\times3} + b \ket{\vec 1}_{3\times3}$. Finally (iii) we  execute a $\mathcal{M}^{(Z)}$ gate in every row, to create the state $\ket{\phi_L}= a\ket{0_L} + b \ket{1_L}$. We can recursively use the same algorithm to create the state at any level of concatenation $k$. Repeating this process recursively yields an error rate for the encoding at the highest level of concatenation $k=L$, $ p^{(L)}_{anc} \leq 10 p^{(0)} + 108 \sum_{j=0}^{L-1} p^{(j)}$. 
\begin{figure}[h]
\subfigure[]{\includegraphics[width=0.4 \columnwidth]{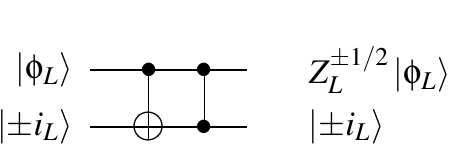}
\label{zhalf}
}
\subfigure[]{
	\centering
		\includegraphics[width=0.4 \columnwidth]{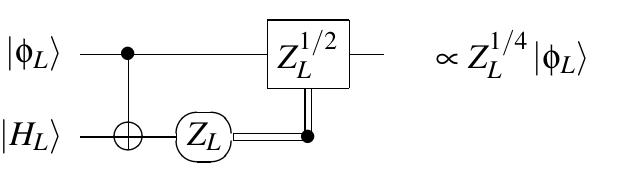}
\label{pi4}	}
	\caption{These circuits need only be implemented at the highest level of concatenation, and thus all operations depicted are encoded operations.  (a) Circuit used to execute an encoded $Z^{1/2}$ Clifford gate on an arbitrary input $\ket{\phi_L}$~\cite{Dennis:02} . (b) Circuit implementing the non-Clifford operation $Z^{1/4}$ given the encoded resource magic state $\ket{H_L}$.}  
\end{figure}
Clearly $p_{(anc)}^{(L)}$ cannot be made arbitrarily small, however, provided $p^{(0)}_{(g)}\leq p_{thresh}$, it can be made small enough to give $p^{(L)}_{(anc)} \leq \sin^2{\pi/8}$, and then one can use MSD to achieve FTUQC~\cite{Bravyi:05}.

Additionally, we promised that preparation errors can in fact be much  
higher than gate error rates. The argument proceeds by using a variant of the algorithmic cooling  
algorithm introduced in Ref. \cite{algcool}.  For a group of three  
qubits $(a,b,c)$ with identical probabilities $p_{(p)}=\epsilon^{(0)}<1/2$, to  
be in the erroneous state $\ket{1}$, we apply $TOFFOLI_{((c,b),a)} CNOT_{(a,c)} CNOT_{(a,b)} 
$. The reduced state of qubit $a$  
is colder, i.e. has lower error ($\epsilon^{(1)}< \epsilon^{(0)})$.   
Concatenating the process, after $j$ rounds using a total of $3^j$ qubits, the final error of the  
one output qubit satisfies the recursion relation $ 
\epsilon^{(j)}=(\epsilon^{(j-1)})^2 (3-2 \epsilon^{(j-1)})$.   
Including gate errors, the total error of this preparation process is  
$p_{(p)}^{(j)}\leq\epsilon^{(j)}+\frac{3}{2}(3^j-1) p_{(g)}^{(0)}$. 

We are now ready to combine our tools. If we are sensibly below threshold, say with $ p_{(g,p)} =0.75 p_{(g,p)thresh} = 2.82 \times 10^{-5}$, then with $  p_{(m)} = 33\%$ we get $  p^{(6)}_{(g)} \sim 10^{-13}$ and $ p^{(6)}_{anc}= 8.32 \times 10^{-3}$ which is safely below the $14.6 \%$ needed for $\ket{H_L}$ distillation (and certainly below the $  50\%$ needed for the $\ket{+i_L}$ distillation~\cite{Aliferis}). { { Thus FTUQC is achievable with noisy and currently achievable measurement error rates, but with only a small impact to the threshold value as compared to the best known result ($1.26 \times 10^{-4}$) for the same code allowing measurements~\cite{Aliferis}.} One can go further and use algorithmic cooling to also push preparation error rates within reach of current technology. We find that {if one has physical preparation error rates of ${p_{(p)}= 1\%}$, then two rounds of AC and physical gate error rates $ p_{(g)} = 2.32 \times 10^{-6}$ allow for FTUQC.} Preparation rates as high as 1/3 can also be allowed, at the cost of demanding a lower gate error rate. For $p_{(p)}\geq1/3$, one can instead use noisy measurement since measurement followed by a unitary is preparation.  

{To put this result in perspective, notice that $p_{(g)} = 1.39 \times 10^{-6}$ is not a threshold value but the required value such that effective preparation and gate error rates are sensibly below our threshold ($0.75 \times p_{thresh}$). In comparison, under the same assumptions the best known result~\cite{Aliferis} implies that quantum computing is possible, with reasonable overhead, when $p_{(p,g,m)} \sim 9.5 \times 10^{-5}$. So the price we pay to push measurement and error rates within reach of current technology (an improvement of three and two orders of magnitude respectively), is demanding roughly two orders of magnitude more stringent gate error rates. The result is even more significant if one considers recent results which show that arbitrarily accurate unitary gates (and not measurement and preparation) can, in principle, be achieved via open system control strategies~\cite{NewLidar}. Furthermore note that the required measurement and preparation error rates have already been reported: in trapped ions~\cite{ITrap}, $p_{(m)} = 2.3\times10^{-3}$ while in quantum dots~\cite{Meunier:06}, $p_{(m)} = 3\times10^{-2}$.}

We point out that the threshold value for gates computed here is by no  
means tight as we wanted to keep calculations simple. We have  
overcounted malignant pairs of locations, and certainly the design of  
our circuits may not be the optimal one in terms of error locations,  
thus in principle the threshold can be improved. On the other hand, restricting ourselves to two-qubit interactions only, and decomposing TOFFOLI gates into one and two qubit  
gates { degrades the gate and preparation threshold value to $2.69 \times 10^{-5}$.} Also restricting to nearest-neighbor only interactions will degrade the threshold value~\cite{Terhal:07}. 
In our circuits ancillas can be prepared offline and we have been careful to limit measurement only to when the data is encoded (at the highest level of concatenation), thus physical systems with slow measurement or preparation are allowed. 

{ In conclusion, we have shown that measurement-free QEC is viable, considerably relaxing the time and error rate constraints on preparation and measurement operations, and pushing them within reach of current technology, while yielding only a small penalty to the gate threshold. This small penalty seems even less relevant if one considers recent results showing that arbitrarily accurate unitary gates can, in principle, be achieved using open system control ~\cite{NewLidar}. Those results complement the methods developed here and bring fault-tolerant quantum computing closer to reality.}

{\it Acknowledgments.-} We acknowledge valuable discussions with P. Aliferis, D. Bacon, and D. Lidar. GAPS acknowledges support from a Macquarie University Research Excellence award and CQCT.

\newpage
\appendix

\section{Details of the gadgets and the threshold calculation}
\label{AppA}
Central to the construction of our error correction routine is the $\mathcal{N}(k)$ gate, which acts as a map between the Bacon Shor (BS) and the Quantum Repetition (QR) code of the same level of concatenation $k$.  It is not evident from the start that $\mathcal{N}(k)$ takes the same amount of time as a gate from our gate library of the same level, nor is it obvious that the failure probability of the corresponding exRec is smaller than that of a CNOT exRec. We show here that indeed this is the case and use these attributes to compute the error threshold. 
\subsection{The $\mathcal{N}$ gate}

\subsubsection{Execution time of $\mathcal{N}$}
One of the main properties we use in our threshold calculation and in our circuits is that $\mathcal{N}(k)$ takes less time than a fully protected gate of the same degree of concatenation $k$, i.e. $T(\mathcal{N} (k)) < T (G(k))$, where $T(A(k))$ denotes the execution time of the $protected$ gate A.  By protected we mean the gate has EC gadgets prepended and appended to the gate. The key observation to prove this is that whenever $\mathcal{N}$ is used, the state only needs protection against one kind of error, for example during the X error correction stage the ancilla only needs protection against X-errors. So to achieve this protection we execute the EC gadget,  $EC_{X}$, without the Z error correction stage. The same analysis follows for the $EC_Z$. In this section we will prove relations explicitly for $\mathcal{N}^{(X)}(k)$, and thus will omit the X or Z superscript, but the reader should have in mind that the same results hold for $\mathcal{N}^{(Z)}$.\\ 

With this in mind, we begin by comparing a fully protected exRec of a gate $G$ at level-$k$ of concatenation with $\mathcal{N}(k)$. The relevant gates can be decomposed as
\beqy
\nonumber G(k) &=& EC(k)\circ \lfloor G(k-1)\rfloor \circ EC(k)\\
\nonumber \mathcal{N}(k) &=& EC_X(k)\circ \lfloor\mathcal{N}(k-1)\rfloor \circ EC_X(k)
\eneqy
where the notation $\lfloor A(k-1) \rfloor $ denotes the implementation of the $A(k)$ gate in terms of $(k-1)$-level protected gates but omitting $k$-level protection. At level $k=1$, $\lfloor A(k-1)\rfloor$ corresponds to the physical gates implementing the encoded gate. Moreover a contracted extended rectangle (exRec) $A(k)$ is composed by the implementation of the gate in terms of level-$(k-1)$ gates and level-$k$ error correcting gadgets in all the inputs and relevant outputs. We use the notation $G(k)$ to denote a fully protected gate made up of a $\emph{single}$ step of level $k-1$ gates, for example a transversal $CNOT(k-1)$ gate.  In contrast the $\mathcal{N}(k)$ consists of more than one step of level-$(k-1)$ protected gates.  To calculate a bound on the time needed perform our error correction, first notice that the full error correction gadget as illustrated in Fig. \eqref{BS} consists of $X$ and $Z$ error correction which both consist of the same number of gates and overlap in all but two locations (neither of which is an $\mathcal{N}(k)$ gate).  Hence, regardless of the structure of $\mathcal{N}(k)$ or it's time duration we find
\beq
\label{ECX-time}
T(EC(k)) = T(EC_X(k)) + 2 T(G(k-1)).
\eneq
Moreover, given the structure of $\mathcal{N}(k)$ (see Fig. \eqref{VN}), it follows then that 
\beqy
\nonumber T(\mathcal{N}(k)) &=& 2 T(EC_X(k))+ 2 T(G(k-1)) + T(\mathcal{N}(k-2))\\
\nonumber T(G(k)) &=& 2 T(EC(k)) + T(G(k-1)) \\
 &=& 2 T(EC_X(k))+ 5 T(G(k-1)),     
\eneqy  
which in turn implies that 
\beq T(G(k)) - T(\mathcal{N}(k)) = 3 T(G(k-1)) - T(\mathcal{N}(k-1)). \eneq 
At level $k=1$ we do know the form and time-duration for all gates in the circuit, and we have that $T(G(1)) - T(\mathcal{N}(1)) = 3 T(G(0)) - T(\mathcal{N}(0)) = 3 T(G(0)) > 0$. So an inductive reasoning leads us to
\beq 
\label{N-time} T(G(k)) - T(\mathcal{N}(k)) > 2 T(G(k-1))>0. 
\eneq
Even more, in the error correction gadgets of level-$k$ we used the subroutine $V\mathcal{N}(k-1)$ which is composed of level $k-1$ gates acting on $9$ level $k-1$ BS encoded inputs and $3$ level $k-1$ QR encoded outputs.  For analysis of this module, the sequence of operations can be decomposed as follows
\beqy 
\nonumber V\mathcal{N}(k-1) &=& EC_X(k-1) \circ \lfloor G(k-1) \rfloor 
\circ EC_X(k-1)  \circ
 \lfloor G(k-1)\rfloor \circ \mathcal{N}(k-1)\\ 
\nonumber &=&  EC_X(k-1) \circ \lfloor G(k-1)\rfloor  \circ EC_X(k-1) \circ \lfloor G(k-1)\rfloor
 \circ EC_X(k-1)  \circ G(k-2) \\
\nonumber && \circ G(k-2) \circ \mathcal{N}(k-2) \circ \mathcal{M}(k-1)\\
\nonumber &=&  EC_X(k-1) \circ \lfloor G(k-1)\rfloor \circ \lfloor G(k-1)\rfloor \circ G(k-2) \circ G(k-2) \circ \mathcal{N}(k-2) \circ \mathcal{M}(k-1)
\eneqy
where in the last equality we have used the exRec-contraction technique from Ref.~\cite{Aliferis}, and used the fact that only the lower output of the gate will be used. The execution time of this contracted exRec satisfies
\beqy 
\nonumber T(V\mathcal{N}(k-1)) &\leq&  2 T(EC_X(k-1))+ 4 T( G(k-2))+ T(\mathcal{N}(k-2))\\
\nonumber &\leq& 2 T(EC_X(k-1))+ 5 T(G(k-2))\\
\nonumber &=& 2 T(EC(k-1))+ T(G(k-2))\\ 
&=& T(G(k-1),   
\eneqy where we have used the fact that $T(EC_X) > T(\mathcal{M})$ to go from line one to two, Eq. \eqref{N-time} from two to three and Eq. \eqref{ECX-time} to get the last equality. This shows why $V\mathcal{N}(k-1)$ takes one level $k-1$ time slot in our circuits.\\

\subsubsection{Error contribution of $\mathcal{N}$ }
Another relevant property for our threshold calculation is the failure probability of a $\mathcal{N}$ gate at some level $k$. For instance, if its failure probability was greater than that of a CNOT of the same level, we would have to equate that into our threshold calculation.  We want to show something even stronger: during $EC(k)$ one uses the subroutine $(V \mathcal{N})(k-1)$ (see Fig. \eqref{VN}) which can be further decomposed as $(V \mathcal{N})(k) =  \prod_{i\in rows} (V \mathcal{N})_i(k)$, we will show that the error probability of the contracted exRec corresponding to the collection of level-$(k-1)$ gates $(V \mathcal{N})_i(k-1)$ is not larger that of a CNOT of level-$(k-1)$. In a more fundamental way we will show that the exRec with the highest failure probability is the one corresponding to the CNOT gate. As in the previous section, we will prove all relations for the $\mathcal{N}^{(X)}$. To simplify notation we drop the X super or sub scripts when necessary but remind the reader that the analysis holds for both X and Z related routines.

The sketch of the calculation is the following. At an arbitrary level of concatenation $k$, the failure probability of an exRec will depend on the failure probability of gates and of $\mathcal{N}$ of lower level-$(k-1)$, but since we do not know {\it ab initio} what is the failure probability of  $(V \mathcal{N})_i(k)$, we only know its failure probability must be larger than the $\mathcal{N}(k)$ one, we cannot directly compute a threshold condition. Fortunately we know what is the specific form of $(V \mathcal{N})_i(1)$ at level $k=1$ in terms of level-0 gates, and we can directly compare it with CNOT(1). In general, this comparison can be done first counting the number of malignant pairs of level-$(k-1)$ errors within a gate $G$ exRec of level-$k$, $A(G(k))$ and written in terms of various malignant-error parameters $\{ A_{EC(k)}, ...\}$ which we will define below. The error probability of such exRec is then given by 
\beq 
p_{G}^{(k)} = A'_{G(k)} (p_{\tilde G}^{(k-1)})^2
\eneq
where $\tilde G$ corresponds to the gate of level $k-1$ with the highest failure probability, $A'_{G(k)} = \frac{A_{G(k)}}{2} \left(1 + \sqrt {1 + \frac{4 B_{G(k)}}{(A_{G(k)})^2}}\right)$, and $B_{G(k)}$ denotes all possible three-site errors in the exRec. Once we show that at level $k=1$, the CNOT is the largest exRec, then we can do the analogue level $k=2$ calculation but now replacing the failure probability of $(V \mathcal{N})_i(k)$ with $p^{(1)}$, now with parameters $\{A_{EC(k=2)},...,\}$. An inductive reasoning will finally lead us to the comparison at any level of concatenation: $p_{CNOT}^{(k)} > p_{(V \mathcal{N})_i}^{(k)}$.

The malignant error parameters are defined as:
\begin{itemize}
\item $A_{EC}$ ($A_{EC_X}$) the number of malignant pairs in an EC ($EC_X$); 
\item $u$ ($u_X$) the number of single failures in an EC ($EC_X$) which  
generate a single error in the data; $\bar u$ ($\bar u_X$  is similar but restricting the error to be in only 8 out of the 9 qubits in the encoded data. This case is important when we have an errors propagating through a CNOT;
\item $\alpha$ ($\alpha_X$) the number of single failures in an EC ($EC_X$) which, in conjunction with an incoming data error, generate a double error in the data; 
\item Parameters $A_{\mathcal{M}}$, $m$, $\bar m$ and $\beta$ can be defined for the quantum  
repetition code and its error correcting gadget, the $\mathcal{M}$ gate. 
\end{itemize}
because at this point we have to assume that the size of the circuits varies with every level of concatenation, then each parameter will have a $(k)$ denoting the level of concatenation it corresponds to. 

Let us now proceed with the calculation. The number of malignant pairs in the $CNOT(1)$, $(V \mathcal{N})_i(1)$, and $\mathcal{N}(1)$ exRecs are then given by
\beqy
\nonumber A_{CNOT(1)} &\leq& (4 A_{EC(1)} + 16 u_{(1)} + u_{(1)} \bar u_{(1)} + 4 u \alpha_{(1)} + 18 \alpha_{(1)} + 36 ) \\
\nonumber A_{(V \mathcal{N})_i(1)} &\leq& (3 A_{EC_X(1)} + A_{M(1)} + 3 u_{X(1)} \bar u_{X(1)} + 66 u_{X(1)}   + 3 u_{X(1)} \beta_{(1)} + 33 \beta_{(1)} + 363),\\
\nonumber A_{bTOFF(1)} &=& (2 A_{EC(1)} + 2 A_{M(1)}) + m_{(1)} \bar m_{(1)} + 2 u_{(1)} \bar m_{(1)} + u_{(1)} \alpha + 2 m_{(1)} \alpha + 8 u_{(1)} + 16m_{(1)} + 9 \alpha_{(1)} + 36
\eneqy
while a direct count gives the value for the malignant-error parameters $\{A_{EC(1)} = 4182, u_{(1)} = 63, \bar{u}_{(1)} = 56 , \alpha_{(1)} = 42 , A_{EC_X(1)} = 2031, u_{X(1)} = 45, \bar{u}_{X(1)} = 30, \alpha_{X(1)} = 20, A_{M(1)} = 177, m_{(1)} = 12, \bar{m}_{(1)} = 8, \beta_{(1)} = 5,\}$. This yields
$p_{(V \mathcal{N})_i}^{(1)} = (11836) (p^{(0)})^2 \leq \frac{1}{2} (33036) (p^{(0)})^2 = \frac{1}{2} p_{CNOT}^{(1)}$ and $p_{bTOFF}^{(1)} = (14784) (p^{(0)})^2 \leq \frac{1}{2} (33036) (p^{(0)})^2 = \frac{1}{2} p_{CNOT}^{(1)}$.  Now, for $k=2$, we obtain the following failure probabilities, using that $bTOFFOLI(1)$ and $(V \mathcal{N})_i(1)$ fail with half the probability of a $CNOT(1)$ gate.

\beqy
\label{kg1}
\nonumber A_{CNOT(k)} &=& (4 A_{EC(k)} + 16 u_{(k)} + u_{(k)} \bar u_{(k)} + 4 u \alpha_{(k)}  + 18 \alpha_{(k)} + 36 ) \\
\nonumber A_{(V \mathcal{N})_i(k)} &=& (3 A_{EC_X(k)} + A_{M(k)} + 3 u_{X(k)} \bar u_{X(k)} + 72 u_{X(k)} + 3 u_{X(k)} \beta_{(k)} + 36 \beta_{(k)} + 432),\\
\nonumber A_{bTOFF(k)} &=& (2 A_{EC(k)} + 2 A_{M(k)}) + m_{(k)} \bar m_{(k)} + 2 u_{(k)} \bar m_{(k)} + u_{(k)} \alpha_{(k)} + 2 m_{(k)} \alpha_{(k)} \\
 && + 8 u_{(k)} + 16 m_{(k)} + 9 \alpha_{(k)} + 36
\eneqy
with corresponding parameter values $\{A_{EC(k>1)} = 1953 u_{(k>1)} = 63,\, \bar{u}_{(k>1)} = 56 ,\, \alpha_{(k>1)} = 33 ,\, A_{EC_X(k>1)} = 1128,\, u_{X(k>1)} = 45,\, \bar{u}_{X(k>1)} = 30,\, \alpha_{X(k>1)} = 16,\, A_{M(k>1)} = 105,\, m_{(k>1)} = 12,\, \bar{m}_{(k>1)} = 8,\, \beta_{(k>1)} = 4,\,\}$. A direct calculation shows again $p_{(V \mathcal{N})_i}^{(2)} < \frac{1}{2} p_{CNOT}^{(2)}$ and $p_{bTOFF}^{(2)} < \frac{1}{2} p_{CNOT}^{(2)}$. From this point on, the structure of the level $k$ error correction circuits, and thus the corresponding malignant error parameter values, are the same of the level $k=2$ circuits, so repeating the process for $k=3,4,...,k$ leads us to the conclusion that the CNOT exRec is in fact the largest exREC to be considered and the one which will determine our threshold value.  

\begin{figure}[h]
     \subfigure[ CNOT(k) exREC ]
{ $\Qcircuit @C=1.0em @R=1.0em {
  & \gate{EC} & \ctrl{1} & \gate{EC}  & \qw  \\
  & \gate{EC} & \targ    & \gate{EC}  & \qw
}$
\label{CNOTEXREC}
}
  \subfigure[ $bTOFFOLI(k)$ exRec ]
{ \label{BTOFFEXREC}
       $\Qcircuit @C=1.0em @R=1.0em {
  & \gate{EC}          & \targ     & \gate{EC}  & \qw  \\
  & \gate{\mathcal{M}_1} & \ctrl{-1} & \gate{\mathcal{M}_2}  & \qw\\
  & \gate{\mathcal{M}_3} & \ctrl{-1} & \gate{\mathcal{M}_4}  & \qw
}$}
\subfigure[ $V\mathcal{N}_{\,i}(k)$ contracted exREC ]
{
$\Qcircuit @C=1.0em @R=1.0em {
  & \gate{{EC_X(k)}} & \gate{W} & \ctrl{2}  &   \\
  & \gate{{EC_X(k)}} & \ctrl{1} &           &  \\
  & \gate{{EC_X(k)}} & \gate{\lfloor X \rfloor}    & \gate{\lfloor X \rfloor}     & \gate{\mathcal{N}^{(X)}{(k-1)}} & \gate{\mathcal{M}^{(X)}(k-1)}
  }$
\label{VNEXREC}} 
  \caption{The largest exRecs to be considered. An EC gate corresponds to a BS QEC routine while a $\mathcal{M}$ gate corresponds to a QR QEC routine. The circuit \eqref{VNEXREC} is executed in every row of the $3\times 3$ array. Because in the $\mathcal{N}$ exRec we are discarding the top-lines we do not require output $\mathcal{M}$ gadgets appended to them. Moreover, at level $k=1$ there is no need for the waiting (W) gate and both CNOTs can be executed simultaneously.}
   \label{ALLEXREC}
   \end{figure}
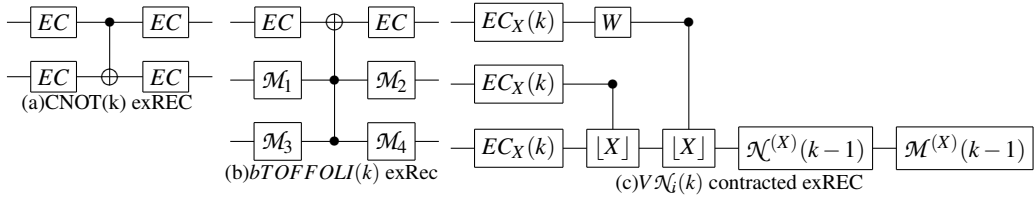

\subsection{Error analysis for the encoder circuit}
The error analysis for the encoder circuit is as follows: to encode a level $k$ state provided a level $k-1$ state, we have that step (i) uses 8 CNOTS, 20 waiting gates, and 8 $\ket{0}$ preparations failing with probability $p^{(k-1)}$ and step (ii) can introduce unwanted phases with a single error (note that this is not a problem in Clifford ancilla preparations e.g. $\ket{0}$ states.) thus we count all locations in the $\mathcal{M}$ gates. A $\mathcal{M}(1)$ contributes with 27 level-0 locations, while a $\mathcal{M}(k)$, for $k>1$, contribute with 24 level-(k-1) locations. So we have
\beq
p^{(L)}_{anc} \leq 10 p^{(0)} + 108 \sum_{j=0}^{L-1} p^{(j)},
\eneq
which justifies our encoder circuit error analysis.

\subsection{Threshold calculation}  
To find the threshold we must now solve the system of equations 
\beqy
\label{recur}
p_{CNOT}^{(1)} &\leq& A'_{CNOT(k=1)} (p^{(0)})^2;\\
\quad p_{CNOT}^{(k)} &\leq& A'_{CNOT(k>1)} (p_{CNOT}^{(k-1)})^2,\textrm{  for  } k>1,
\eneqy 
which in turn gives that, $p_{(p,g)thresh} =  \frac{1}{\sqrt{A'_{CNOT(k=1)} A'_{CNOT(k>1)}}} = 3.76 \times 10^{-5}$. Because no measurement exRec of lower levels had to be used in our calculations, this threshold only applies to gate and preparation. To find what is the Clifford measurement error rate threshold , we note that at the highest level of concatenation
\beqy
\label{meas} p_{(m)}^{(k+1)} &\leq& A_{EC(k)}(p^{(k)})^2 + 2 u_{(k)} p^{(k)} p_{(m)}^{(k)}+ 3 (p_{(m)}^{(k)})^2. 
\eneqy
So if preparation and gate error rates are below threshold, then for $k$ large enough $p^{(k)}$ is vanishingly small and the first two terms can be neglected. Then the threshold condition for $X$ and $Z$ measurements is $ p_{(m)thresh} = 1/3$. \\

\subsubsection{Threshold calculation with two-qubit interactions only}
In our library of gates we assumed the possibility of executing 3-qubit gates in the form of a TOFFOLI gate. Threshold computations using this gate are standard, see e.g.\cite{Aharonov:99}. Although for some architectures it is reasonable to include three qubit gates in the gate library~\cite{threequbits}, but for others two qubit interactions are mor e natural. Here we show how to compute the threshold when decomposing TOFFOLI gates into one and two qubit gates. The main observation is that this decomposition only affects the level-1 of the concatenation level, where we actually replace the TOFFOLI as shown in Fig. \eqref{extTOFF}
\begin{figure}[htbp]
	\centering
		\includegraphics{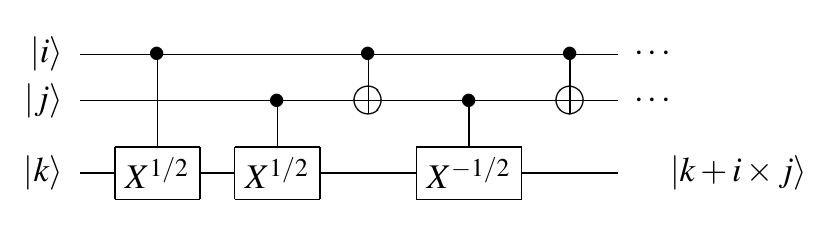}
	\caption{Decomposition of a physical TOFFOLI gate into two qubit interactions. Note that it requires only three time-steps as the first two $CX^{1/2}$s can be executed simultaneously. The last CNOT gate is not necessary as we will typically discard the controls of such TOFFOLI and thus they do not count towards our threshold estimation. }
	\label{extTOFF}
\end{figure}
Lets recall the level-1 bTOFFOLI for the BS code is composed by 3 physical TOFFOLI targeting one row, and controlled by three classical three bit strings, and 6 waiting gates. Since a physical TOFFOLI can be executed in three time steps and is composed of five gates, then the encoded bTOFFOLI gate is composed of $3 \times 6 + 3 \times 5 = 33$ physical gates. Note that this decomposition mainly affects level-1 of concatenation, as we are only decomposing the physical gate and not emulating such decomposition at all levels of concatenation. If we input this in our analysis, an analogue hierarchy of exRecs is maintained  i.e. $ p_{bTOFF(k)} \leq \frac{1}{2} p_{CNOT(k)}$ and $p_{(V \mathcal{N})_i(k)} \leq \frac{2}{3} p_{CNOT(k)}$, which in turn yields a threshold of $p_{(p,g) thresh} = 2.68 \times 10^{-5}$ now limited to two-qubit gates only.

\section{Tools for other CSS codes}
\label{AppB}
At a first glance the tools developed here are specialized for the $3\times 3$ BS code, however we want to show that they can also be adapted to other CSS codes. We present here some tools which will be necessary to generalize our methods.\\  

\subsection{$\mathcal{M}$ gate for larger codes}
If we were to use larger codes, e.g. the $25$ qubit BS code, one would typically need to execute larger majority votings, i.e. of a longer distance. To achieve this purpose we developed a way of executing majority voting in a fault-tolerant way or, equivalently, a way of fault-tolerantly and unitarily correcting a quantum repetition code. We denote this gate as $\mathcal{M}(N)$.
\beq
\mathcal{M}(N): \ket{ s_1,s_2,...,s_{N}}_{N}\otimes{\ket{\epsilon_0}} \rightarrow \ket{m,...,m}_{N}\otimes{\ket{\epsilon_{\{s_i\}}}} = \ket{\vec{m}},
\eneq
where $m = MBF\{s_1,s_2,...,s_{N}\}$, $\ket{\epsilon_0} = \ket{\vec{0}}^{\otimes N}$ and MBF is the majority boolean function. Note that when $N=2k$ i.e. is even, the MBF may not be solvable, i.e. when the string is balanced, in those cases the protocol will just take $\{s_1,s_2,...,s_{2k}\}$ to another balanced $\{s'_1,s'_2,...,s'_{2k}\}$. 

Consider the $\mathcal{M}$ gate in Fig. \ref{fig:classical}.
\begin{figure}[h]
\beq
\nonumber \Qcircuit @C=1.0em @R=.7em { 
{\ket{\vec{s}}} & & \ctrl{6} & \qw & \ctrl{6} & \multigate{6}{\mathcal{C}}& \qw  & \rstick{\ket{\vec{s}'}}\\
{\ket{\vec{0}}}&& \targ & \gate{R} & \targ & \ghost{\mathcal{C}} & \qw & \cdots\\
{\ket{\vec{0}}}&& \targ & \gate{R^2}& \targ & \ghost{\mathcal{C}} & \qw & \cdots \\
{\ket{\vec{0}}}&& \targ & \gate{R^3} & \targ & \ghost{\mathcal{C}}& \qw & \cdots\\
{\ket{\vec{0}}}&& \targ & \gate{...} & \targ & \ghost{\mathcal{C}}& \qw & \cdots\\
{\ket{\vec{0}}}&& \targ & \gate{R^{N-1}} & \targ & \ghost{\mathcal{C}}& \qw & \cdots\\
{\ket{\vec{0}}}&& \targ & \gate{R^N} & \targ & \ghost{\mathcal{C}}& \qw & \cdots \gategroup{1}{3}{7}{5}{.7em}{--}
}
\eneq	
\caption{The boxed part of the circuit is in charge of the syndrome extraction, $R$ corresponds to a cyclic permutation of the physical qubits and $\ket{\vec{s}}$ denotes a codeword $\ket{s_1,...,s_N}$. The last unitary $\mathcal{C}$, targeting the top line, will be the one in charge of executing the desired operation, depending of how we choose the controls as we will see below. All operations depicted here are bitwise and thus transversal.}
\label{fig:classical}
\end{figure}
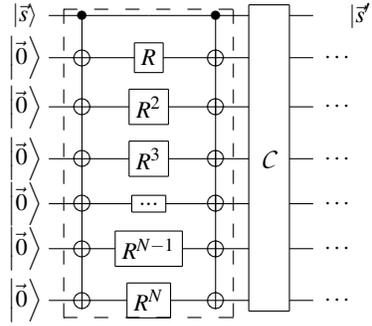
The $\mathcal{C}$ gate can now be chosen to be a series of multi-controlled NOT gates targeting the data string. In general, it will be all possible $k$-controlled-NOT gates targeting the data, where $k$ is to be chosen from a set $K_N$ which is characteristic for every string length $N$, e.g. $ K_3=\{ 2\}, K_5 =\{4,3 \}, K_7 =\{6,4\}$. For the purposes of this paper the $N=3$ case is of special interest and $\mathcal{C}$ is just a TOFFOLI gate which is assumed in our library of $physical$ operations. For notation purposes we write $\mathcal{M}$ for  $\mathcal{M}(3)$. Note that with this majority voting gadget we can build the corresponding EC gadget for larger BS codes, in the same way we used the $\mathcal{M}$ to build the $3\times 3$ BS code QEC routine. \\
\subsection{Parity voter ($\bf \mathcal{P}$)} 
Other circuit which we do not use here, but may be of use is the {\it Parity voter} circuit, $\mathcal{P}$:
\beq
\mathcal{P}(N): \ket{ s_1,s_2,...,s_{N}}_{N}\otimes{\ket{\epsilon_0}} \rightarrow \ket{q,...,q}_{N}\otimes{\ket{\epsilon_{\{s_i\}}}} = \ket{\vec{q}},
\eneq
where $q = s_1 \oplus s_2 \oplus ... \oplus s_{N}$. This gate can be executed slightly modifying the circuit in Fig. \eqref{fig:classical}: we (i) swap the order of the $\mathcal{C}$ and the last set of CNOTs and, (ii) define $\mathcal{C}$ as all the possible CNOTs controlled by the ancilla qubits in a column and targeting the data string qubit corresponding to that column. Let us note that we can also vote the parity of a string $\ket{q_1,q_2,...,q_S}$ into it's last qubit via $\prod_{k=1}^S CNOT_{i,k}$.
\\ 
\subsection{$\ket{Cat}$ state verification}
In contrast to the 9-qubit Bacon-Shor code other codes need an extra element: $\ket{cat}$ state verification. This verification stage in general provides a test which if passed gives an outpput state which when used an ancilla in the QEC process will not ruin the fault-tolerance, and if failed indicates that the whole ancilla preparation \& verification process must be restarted. The method we develop here will not be a ``test'' but rather a $deterministic$ way of producing ``verified'' output $\ket{cat}$-states. 
    
Depending on the EC gadget of choice one will typically need a way of verifying the state 
\beq
\ket{cat} = (1/\sqrt{2}) (\ket{0000} + \ket{1111}).
\eneq
Typically these states are prepared and verified through a measurement: if the verifier qubit has not flipped then the state passes the test, if it has flipped then the preparation \& verification process must be restarted. The whole idea is that a two bit-flip cat-state such as $(\ket{0011} + \ket{1100})/\sqrt{2}$ has to pass the verification with probability $p^2$ or worse, such that the fault-tolerance is maintained. 

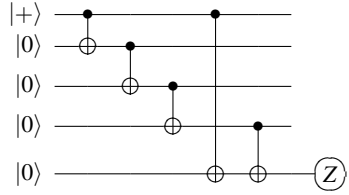
\begin{figure}[h]
\beq \nonumber\Qcircuit @C=1.0em @R=1.0em { 
\lstick{\ket{+}}   & \ctrl{1} & \qw       & \qw     & \ctrl{4} &  \qw & \qw & \\
\lstick{\ket{0}}   & \targ    & \ctrl{1}  & \qw     &  \qw    &  \qw & \qw & \\
\lstick{\ket{0}}   & \qw      & \targ     & \ctrl{1}&  \qw    &  \qw & \qw & \\
\lstick{\ket{0}}   & \qw      & \qw       & \targ   &  \qw    &  \ctrl{1} & \qw & \\
\lstick{\ket{0}}   & \qw      & \qw       &  \qw    &  \targ  &  \targ   & \qw & \measure{Z}
}
\eneq
\caption{Shor's measurement aided $\ket{cat}$ preparation and verification. Two errors, a lethal scenario for quantum error correction, show up in the final state only when two or more operations are faulty.}
\end{figure}
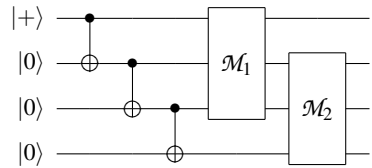
\begin{figure}
\beq \nonumber \Qcircuit @C=1.0em @R=1.0em { 
\lstick{\ket{+}}   & \ctrl{1} & \qw       & \qw     & \multigate{2}{\mathcal{M}_1} &  \qw               &  \qw & \\
\lstick{\ket{0}}   & \targ    & \ctrl{1}  & \qw     &  \ghost{\mathcal{M}_1}    &  \multigate{2}{\mathcal{M}_2} & \qw & \\
\lstick{\ket{0}}   & \qw      & \targ     & \ctrl{1}&  \ghost{\mathcal{M}_1}    &  \ghost{\mathcal{M}_2} & \qw & \\
\lstick{\ket{0}}   & \qw      & \qw       & \targ   &  \qw                    &  \ghost{\mathcal{M}_2} & \qw & 
}
\eneq
\caption{We replace the verification measurement by two successive actions of $\mathcal{M}$, the circuit will output a $bad$ cat if there are two failures within the circuit, i.e. a $p^2$, or higher order event in $p$, as desired. The same method can be executed for larger cats still admitting only one error, using longer $\mathcal{M}$ cascades. For cat states admitting more errors we just use cascades of larger versions of $\mathcal{M}$. }
\label{Mver}
\end{figure}
Note that this state is the result of a single failure of a CNOT during the preparation stage, however the measurement gives us a criterion for discarding it. An extra failure in the measurement must happen for the $bad$ cat state to pass the test, but that is already a $p^2$ event so the analysis for fault-tolerance is valid. This process is non-deterministic in the sense that one error in the measurement can lead to the rejection of a perfectly good ancilla. As we want to avoid measurements we would ideally want to avoid such process altogether, thus we can execute the circuit in Fig. \ref{Mver}.
This implies that we have a way of unitarily and deterministically, i.e. no discarding and restarting the process, preparing our $verified\ $$ \ket{cat}$ state ancilla.\\

\subsection{$\mathcal{M} \& \mathcal{N}$ for other CSS codes} 
Beyond the circuits that we presented here, in a deeper sense what our $\mathcal{N}^{(X)}$ ($\mathcal{N}^{(Z)}$) circuit does is check the parity of all representations of logical X (Z) operators modulo stabilizer operators, and project this information into a X (Z) QR code. For example, the Steane code is defined by the stabilizers $\{ IIIXXXX, IXXIXX, XIXIXIX, IIIZZZZ, IZZIZZ, ZIZIZIZ \}$, so there are seven different implementations of $X_L$ ($Z_L$) using only three X (Z) physical gates. So to execute the $\mathcal{N}^{(X)}$ gate we execute the following protocol for all the seven logical operators. For the operator $X_{Z_i}$, obtained by applying X operators on the qubits $\alpha_i,\beta_i,\gamma_i$ of the codeword we:  (i) prepare an ancilla $\ket{000}$ state; (ii) a bitwise CNOT between the qubits $\alpha_i,\beta_i,\gamma_i$ of the codeword and the ancilla, the resulting state of the ancilla is a string $\ket{s_i}$; (iii) vote the parity of $\ket{s_i}$ into its last qubit, $\ket{p(s_i)}$. Note that the process can be executed simultaneously, because no qubit is targeted twice and we are not concerned by propagation of $Z$-errors into the data. Finally we execute an $\mathcal{M}_7$ gate on the string to obtain the final state. So we have achieved the mapping 
\beq
\mathcal{N}: \ket{x_L} \rightarrow \mathcal{M}(\bigotimes^7_{i=1}\ket{p(s_i)}) = \ket{\vec{x}}
\eneq
This process is fault-tolerant and needs two errors for the gate to fail. To see this, note that one error in the data, or in (i) to (iii) may at most generate a string with three errors which can still be corrected by $\mathcal{M}$, e.g. $\ket{0_L}\underrightarrow{\textrm{   1 error   }} \ket{1110000} \underrightarrow{\,\,\,\mathcal{M}\,\,\,} \ket{\vec{0}}$. An extra error during the simultaneous (i)-(iii) or in the $\mathcal{M}$ will generate the wrong execution of the $\mathcal{N}$. 

This completes the basic tools needed to adapt our scheme to other CSS codes. Since unitary QEC circuits already exist in the literature for other CSS codes~\cite{Aharonov:99, Ngate} using TOFFOLI gates, one now has to simply replace the TOFFOLI by bTOFFOLI to execute them fault-tolerantly at any level of concatenation. Note that the results regarding the timing on a $\mathcal{N}(k)$ gate still hold for this case since (ii)+(iii) take only $2 \times T(G(k-1))$, it is also true that $T(EC(k)) = T(EC_{X}(k)) + d T_(G(k-1))$ for some integer $d$, which will allow the same analysis as for the BS code. The form of $\mathcal{N}$ can be modified for a particular QEC circuit in order to optimize the error count.


\begin{thebibliography}{99}
\bibitem{fn01}
The Clifford group on a set $V$ of qubits is $\langle \{H_i,CNOT_{i,j}, Z_i^{1/2}\ \forall i,j\in V\}\rangle$.
\end{thebibliography}
\end{document}